\documentclass[12pt]{iopart}
\usepackage{graphicx}
\usepackage{amssymb}
\begin{document}

\title{Efficiency of Dopant-Induced Ignition of Helium Nanoplasmas}

\author{A. Heidenreich$^{1,2}$, B. Gr{\"u}ner$^{3}$, M. Rometsch$^{3}$,
S. R. Krishnan$^{4}$, F. Stienkemeier$^{3}$, and M. Mudrich$^{3}$}

\address{$^1$Kimika Fakultatea, Euskal Herriko Unibertsitatea (UPV/EHU)
and Donostia International Physics Center (DIPC), P.K. 1072, 20080
Donostia, Spain\\
$^2$IKERBASQUE, Basque Foundation for Science, 48011 Bilbao, Spain\\
$^3$Physikalisches Institut, Universit{\"a}t Freiburg, 79104 Freiburg,
Germany\\
$^4$Department of Physics, Indian Institute of Technology - Madras, Chennai 600036, India}

\begin{abstract}
Helium nanodroplets irradiated by intense near-infrared laser pulses ignite and form
highly ionized nanoplasmas even at laser intensities where helium
is not directly ionized by the optical field, provided the
droplets contain a few dopant atoms. We present a combined theoretical
and experimental study of the He nanoplasma ignition dynamics for
various dopant species. We find that the efficiency of dopants
to ignite a nanoplasma in helium droplets strongly varies and mostly depends on (i) the pick-up
process, (ii) the number of free electrons each dopant donates upon
ionization, and remarkably, (iii) by the hitherto unexplored effect
of the dopant location in or on the droplet.
\end{abstract}


\section{Introduction}
Plasmas formed in nanoscale matter by the interaction with intense light pulses ranging from near-infrared (NIR) up to hard X-rays are a focus of current research. These studies are motivated by a large number of potential applications including the generation of energetic electrons and ions~\cite{Saalmann:2006,Fennel:2010} as well as intense XUV and attosecond pulses~\cite{Stebbings:2011}. Besides, in single-shot
X-ray imaging experiments of large molecules~\cite{Neutze:2000} and clusters~\cite{Gorkhover:2012}, the creation of an expanding nanoplasma generally causes severe limitations to the achievable resolution of the initial structure. Controlling the nanoplasma dynamics for the purpose of exploiting its exceptional properties or for mitigating its detrimental effects requires in both cases a profound understanding of the dynamics of the ignition and the evolution of such nanoscale atomic and molecular systems in intense light fields. In the NIR excitation regime, the remarkable properties of nanoplasmas have been rationalized by a resonant interaction between the external light field and the dipolar oscillations in the collective electron motion driven by this field~\cite{Saalmann:2006,Fennel:2010}. The resulting efficient light absorption induces avalanche-like charging and heating of the nanoplasma followed by hydrodynamic expansion and Coulomb explosion.

Atomic-scale design of nanoscopic targets provides an attractive
route to controlling the ionization processes herein. Thus, even random
placement of dopants into host clusters has resulted in significant
enhancement of photon or electron emission~\cite{Jha05,Jha10}.
Much better defined two-component nanometer-sized systems are obtained
by embedding atoms as dopants into superfluid helium (He) nanodroplets. Novel
resonant dynamics have been uncovered for these systems when driven
by intense NIR laser pulses~\cite{dodi07,DoeppnerPRL:2010,Peltz:2011,Goede:2013,Mikaberidze:2008,MikaberidzePRL:2009,KrishnanPRL:2011,Krishnan:2012,Krishnan:2014,Mudrich:2014}.
These reports have triggered a search for the optimal conditions for
doping in such clusters. 

He nanodroplets are ideal candidates for probing the effects of dopants properties on the ionization dynamics due to their extreme inertness with respect to chemical reactivity
and to NIR radiation, their homogeneous superfluid density distribution,
and the simple electronic structure of constituent He atoms. Furthermore,
they offer the unique opportunity to vary the location of dopant atoms
by appropriately choosing the atomic species~\cite{Toennies:2004,Barranco:2006}.
While most dopants such as rare gases submerge into the droplet interior,
alkali metal atoms and small clusters reside in dimple-like states
at the droplet surface~\cite{Stienkemeier2:1995,AncilottoAlkalis:1995,Stienkemeier:2006}.
Alkaline earth metals represent an intermediate case in that they generally
localize within the surface layer~\cite{Barranco:2006,Hernando:2007}. A particularly curious dopant system is magnesium atoms which aggregate as a metastable foam-like structure in He droplets~\cite{Przystawik:2008}. By simultaneously doping different species the location of the individual components can be varied due to the pulling of the submerged dopant to the surface~\cite{Douberly:2007} or vice versa~\cite{Lugovoj:2000}. For instance, by co-doping He droplets with calcium (Ca) and xenon (Xe) it was shown that Ca atoms are attracted towards the Xe cluster residing at the droplet center~\cite{Lugovoj:2000}. For the case of mixed doping with rubidium (Rb) and Xe atoms, the dopants are expected to remain separated one from another due to the presence of a potential energy barrier~\cite{Poms:2012}.

Our investigations exploit the singular feature of He nanodroplets to localize dopants in the interior or at the surface and we demonstrate that the dopants' location plays a crucial role in the ionization by intense NIR pulses. We show both in experiment and atomistic molecular dynamics (MD) simulations, that the capability of dopants to trigger avalanche-like ionization of the He host droplet leading to nanoplasma formation in a regime where the laser pulse does not directly ionize the host matrix -- which we refer to as \emph{ignition} -- conspicuously varies as function of the dopants physico-chemical properties. 
The efficiency of dopant-induced ignition is assessed by comparing various species -- Xe residing in the droplet interior, Ca in the surface layer, and potassium (K) on the surface. A detailed analysis reveals the crucial properties of dopants which allows us to propose principles for optimally designing dopant clusters for nanoplasma studies.

\section{\label{sec:experiment}Experiment}
The experimental setup is composed of an amplified fs laser system
for generating intense NIR laser pulses and a molecular beam apparatus
for generating doped He nanodroplets. The latter has been described
in detail elsewhere~\cite{Mudrich:2009,Krishnan:2012}. 

In short, a beam of He nanodroplets is generated by continuously expanding pressurized He ($p_{0}=50$ bar) of high purity (He 6.0) out of a cold nozzle ($T_{0}=18$ K) with a diameter of $5$ $\mu$m into vacuum. At these expansion
conditions, the mean droplet size is $\langle N\rangle\approx5000$
He atoms per droplet~\cite{Toennies:2004,Stienkemeier:2006}. The
He droplets are doped with rare gas and metal atoms by passing through
a scattering cell which contains atomic vapor with adjustable pressure.
When increasing the doping pressure of either rare gas or metal vapor
the He droplets pick up few atoms which aggregate to form clusters
inside (rare gases) or at the droplet surface (alkali, alkaline earth
metals)~\cite{Lewerenz:1995}. At higher vapor pressures, the He
droplets undergo massive scattering and shrinkage due to evaporation induced by the transfer of transverse momentum and the deposition of kinetic and
binding energy (dopant-dopant and dopant-He). This leads to a general
decrease in all droplet-correlated signals. The average number
of dopants attached to the He droplets is inferred from the measured
dopant partial pressure using detailed simulations of the pick-up
process~\cite{Buenermann:2011}. In the detector chamber, the doped
He droplet beam crosses the focused laser beam at right angles in
the center of a standard time-of-flight (tof) ion mass spectrometer.

The laser pulses (center wave length $\lambda=800$ nm, pulse length
$t_{\mathrm{FWHM}}=220$ fs) are generated by chirped pulse amplification (Coherent Legend) at a repetition rate of 5 kHz. The pulses are focused by a lens (focal length $f=75$ mm) placed inside the detector chamber to reach a maximum peak intensity $I=5\times10^{15}\thinspace\mathrm{W\, cm^{-2}}$ in the focal volume.

\section{\label{sec:theory}Theory}

The MD simulation method for the interaction of a cluster with the electric and magnetic field of a linearly polarized NIR Gaussian laser pulse was described previously~\cite{Heidenreich07,HeidenreichIsrael07,Heidenreich:2012}. All atoms and nanoplasma electrons are treated classically, starting with a cluster of neutral atoms. Electrons enter the simulation, when the criterion for tunnel ionization (TI), classical barrier suppression ionization (BSI) or electron impact ionization (EII) is met. This is checked at each atom at every MD time step, using the local electric field at the atoms as the sum of the external laser electric field and the contributions from all other ions and electrons of the cluster. Instantaneous TI probabilities are calculated by the Ammosov-Delone-Krainov (ADK) formula~\cite{Ammosov:1986}, EII cross sections by the Lotz formula~\cite{Lotz:1967} taking the ionization energy with respect to the atomic Coulomb barrier in the cluster~\cite{Fennel:2007}.
The effect of chemical bonding on the valence shell ionization energies of K and Ca dopants is disregarded. 

Coulomb potentials between ions, smoothed Coulomb potentials for ion-electron and electron-electron interactions are used. Interactions involving neutral atoms are disregarded except for a Pauli repulsive potential of 1.1 eV between electrons and neutral He atoms~\cite{Buchenau:1991} in terms of a fourth-order Gaussian function centered at every He atom. The binding potentials of He$_2^+$ and of other He$_n^+$ complexes are not implemented, so that the simulations cannot account for the He$_2^+$ formation explicitly; we can only estimate an upper bound of the He$_2^+$ abundance from the remaining groundstate neutral He atoms and He$^+$ ions at the end of each trajectory. Neutral He atoms and He$^+$ ions which are formed by three-body electron-ion recombination are Rydberg state atoms and are therefore excluded from the estimate of the He$_2^+$ production. Electron-ion pairs which are found within a cutoff distance of 2\,\AA~at the end of each trajectory (temporal length 0.7-1.8~ps) are taken to be recombined and the ion charge state abundances are corrected accordingly. 

He ion and dopant ion signals are laser-intensity averaged over the three-dimensional focus volume~\cite{Heidenreich:2011} in the range $8\times 10^{12}$-$5\times 10^{15}$~Wcm$^{-2}$. Due to the high sensitivity of the droplet evolution to initial conditions, the results are averaged over sets of 5 to 100 trajectories per doped droplet and laser intensity. Moreover, surface-doped droplets (K and Ca) are averaged over their parallel and perpendicular orientations of the dopant-droplet axis with respect to the laser polarization axis unless mentioned explicitly. The temporal width of the Gaussian pulse intensity envelope is $\tau_I=200$~fs, slightly lower than in the experiment (220~fs).

\begin{figure}[hbt]
\center
\includegraphics[width=0.6\columnwidth]{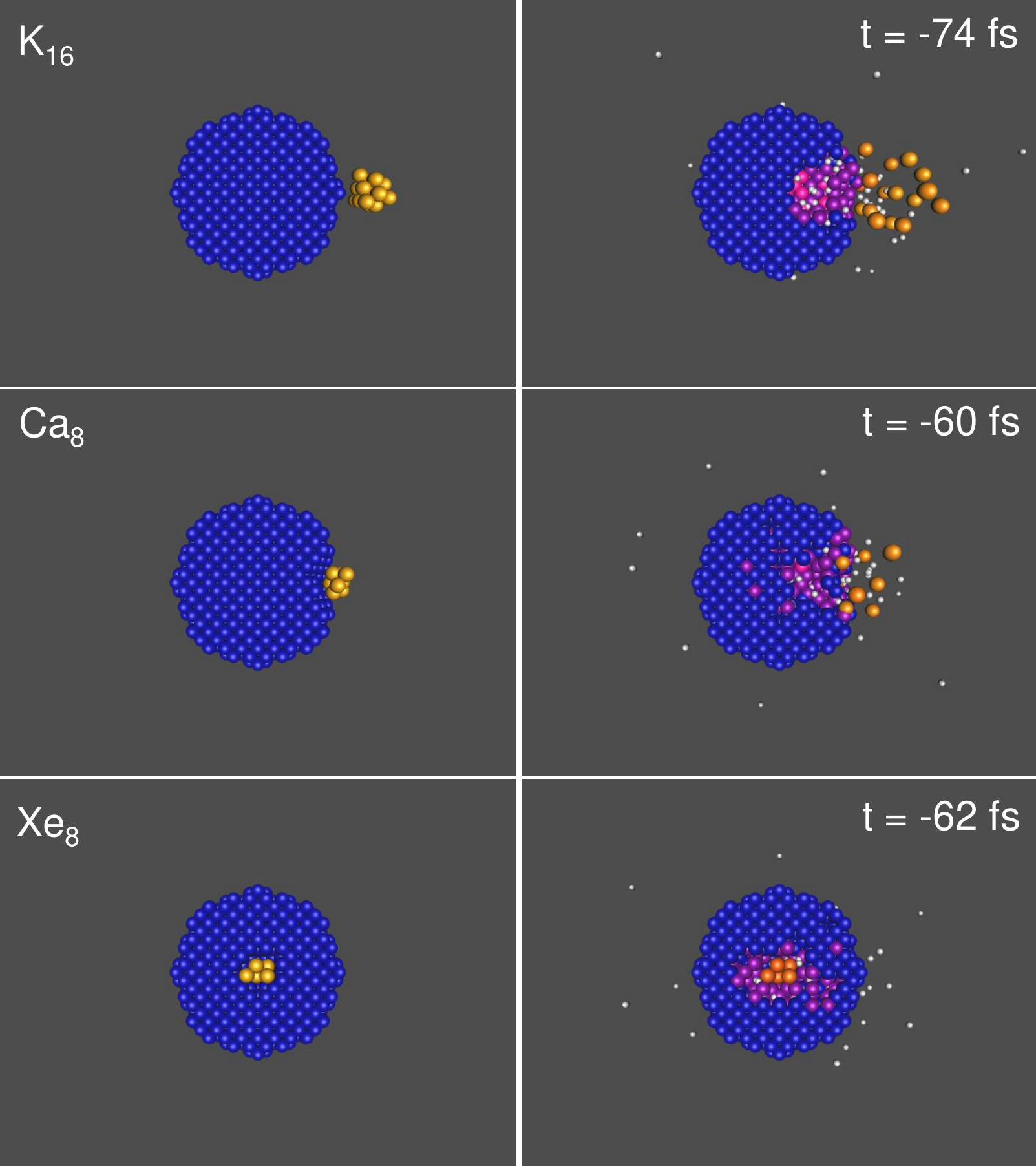}\caption{\label{fig:snapshots} Cross sectional views of the doped He droplets before (left column) and at the onset of droplet ignition $60-74$~fs before the maximum of the laser pulse (right column). The blue, orange, purple and white bullets depict neutral He atoms, dopant atoms, ions, and electrons, respectively.}
\end{figure}

For the He droplets we assume a fcc structure with an interatomic distance of 3.6\,\AA~\cite{Peterka:2007}. The dopant clusters are assembled according to the principle of densest packing of tetrahedra and to form, as far as possible, spherical shapes. We use the following interatomic distances: K-K 4.56\,\AA~(taken as the average interatomic distance in a K$_{20}$ cluster)~\cite{Banerjee:2008}, Ca-Ca 3.9\,\AA~ (average value for Ca clusters)~\cite{Mirick:2001}, Ca-Xe 5.17\,\AA~ (CaXe complex)~\cite{Czuchaj:2003}, Xe-Xe 4.33\,\AA~(bulk), He-Xe 4.15\,\AA~\cite{Chen:1973}, He-K 7.13\,\AA~\cite{Hauser:2013}, He-Ca 5.9\,\AA~(HeCa diatomic complex)~\cite{Hinde:2003}. 

In case of surface doping we assume a dimple depth of 7~\AA~(inferred from density functional calculations of a single Ca atom on the surface of a He droplet)~\cite{Hernando:2007}. According to Ancilotto et al.~\cite{AncilottoAlkalis:1995}, a single K atom is located in a dimple of depth 2.3\,\AA. Since such a shallow dimple cannot be implemented in a fcc lattice of discrete He atoms, we neglect the dimple for K dopants. The left column of Fig.~\ref{fig:snapshots} shows cross sectional views of dopant-He complexes containing 2171 He atoms and the indicated number of dopant atoms. The right column shows snapshots of the clusters shortly after He ignition at the indicated interaction times with respect to the maximum of the laser pulse of peak intensity $I=10^{14}\mathrm{\thinspace W\, cm^{-2}}$.  Neutral He atoms are represented by blue spheres, dopant atoms are orange, ions are red and electrons are small white dots.

\begin{figure}[hbt]
\center
\includegraphics[width=0.65\columnwidth]{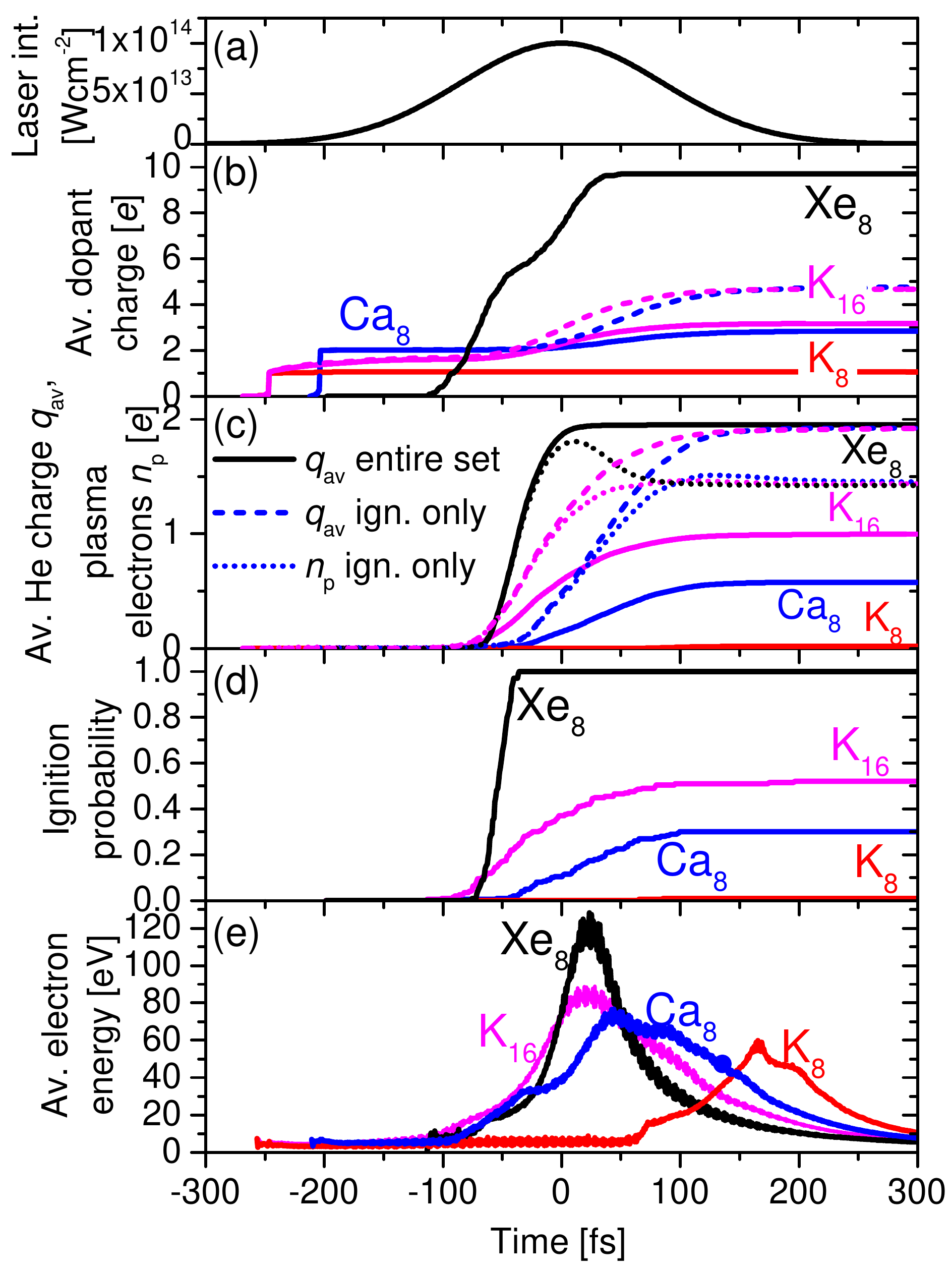} \caption{\label{fig:Charging} Simulated ionization dynamics of a He$_{2171}$ droplet doped with Xe$_8$, K$_8$, K$_{16}$ and Ca$_8$ clusters for fixed pulse peak intensity $I = 10^{14}$ Wcm$^{-2}$. (a) Temporal profile of the Gaussian pulse intensity envelope; (b) Average charge
per atom of dopants and (c) of He atoms. (d) Ignition probabilities; (e) Average electron kinetic energies. All quantities are trajectory set-averaged.}
\end{figure}

\section{\label{sec:Results}Results}
Our simulations provide full insight into the nanoplasma ignition and charging dynamics by giving access to all relevant microscopic and ensemble-averaged observables as they evolve in time, including electron and ion kinetic energies as well as the yields and charge states of He and dopant ions. As examples, Fig.~\ref{fig:Charging} illustrates for selected dopant species the dynamics of ionization of dopants and of the He host initiated by the ignition and avalanche-like growth of a He nanoplasma. 
Shown are the intensity envelope of the laser pulse in panel a), the average charge per dopant atom b) and per He atom c), the probability of igniting a He nanoplasma d), and the average electron kinetic energy e), for trajectory bundles of He$_{2171}$ droplets doped with clusters K$_8$ and K$_{16}$ (both on the surface), Ca$_8$ (in a dimple on the surface), as well as Xe$_8$ (in the center). 

Ionization of the doped droplets starts with TI or BSI of the dopant in the rising edge of the laser pulse. After a time delay which we call ``incubation time'', ignition of the He droplet induced by EII occurs. The role of the dopant is to provide the seed electrons and to assist EII by lowering of the Coulomb barrier at He by the field of the dopant cations. EII is also the by far dominating ionization channel (typically $>95$\%) in the subsequent rapid charging up as a result of the EII avalanche ionization~\cite{Saalmann:2006,Fennel:2010}. The surprising result at first sight is that both the dopants and the He host atoms charge up to the by far highest charge states for the Xe$_8$ case as compared to K doping (9.7 and 2, respectively, in units of the elementary charge $e$), in spite of the much higher first ionization energy of Xe (12.1 eV) than for K (4.3 eV).

For K dopants, a single ionization per K atom occurs early in the pulse at $t\approx -250$ fs [Fig.~\ref{fig:Charging} b)]. For K$_{16}$, an incubation time of $100$-$150$ fs elapses until the average charge of He [Fig.~\ref{fig:Charging} c)] starts to rise. During this incubation time, EII of He competes with a partial drain of the seed electrons by outer ionization~\cite{Last:1999}. This competition is not always in favor of EII. It turns out that, depending on slight variations of the trajectories' initial conditions but for the same pulse parameters, either He ionization does not take place at all, ceases after a few He atoms, or ignition occurs, that is, ionization propagates avalanche-like through a large part or the entire He droplet. Which factors contribute to ignition and determine the incubation time, is subject of a detailed mechanistic study and will be published in a subsequent paper. We define an average He charge of 0.1 as an empirical criterion for the detection of the onset of ignition. The exact choice of the threshold value is uncritical in view of the rapid charging process in the He droplet. 

The sensitivity to the initial conditions is high for those dopant sizes and pulse parameters for which the occurrence of ignition is on the knife's edge. In those trajectories with ignition, the average charge per He atom jumps to almost 2 within $50$-$70$ fs, and the trajectory set-averaged He charge jumps to 1.0 [Fig.~\ref{fig:Charging} c)] with an ignition probability of 0.52. For these pulse parameters ion-electron recombination and He$_2^+$ formation do not occur; recombination becomes important for $I\lesssim 2\times 10^{13}$ Wcm$^{-2}$. Averaging only over those trajectories with ignition also leads to a long-time average He charge of nearly 2 [included as dashed lines for K$_{16}$ and Ca$_8$ in Fig.~\ref{fig:Charging} c)]. The ignition probabilities, derived as fractions of the number of trajectories with ignition, are shown in Fig.~\ref{fig:Charging} d). The long-time ignition probabilities for K$_{8}$, Ca$_8$ and Xe$_8$ dopants are 0.01, 0.30 and 1, respectively.

The K$_{8}$ and K$_{16}$ examples reveal a dopant cluster size effect: The ignition probability increases with the number of dopant atoms as a larger dopant cluster provides more seed electrons and a stronger electric field created by the sum of dopant ion charges, thereby assisting EII by reducing the Coulomb barrier at the adjacent He atoms. Moreover, the dopant cluster size effect is nonlinear: While for K$_{8}$ the average charge per K atom remains 1 during the incubation time [Fig.~\ref{fig:Charging} b)], for K$_{16}$ it rises gradually, making via EII also up to two inner shell electrons available for seed ionizations. Ca$_{8}$ contributes by its two valence electrons per atom.

With an ignition probability of 1, the Xe$_{8}$ cluster has the highest ignition efficiency as compared to the K and Ca dopants of the same size. For Xe$_{8}$, the initial seed ionization begins much closer to the center of the laser pulse ($t\approx -100$ fs) than for K and Ca dopants. Within an incubation time of only a few fs, up to three seed electrons per atom are set free by TI, BSI and EII, in contrast to only one seed electron per atom in the K$_{8}$ case. This can be rationalized by the relatively low second and third ionization energies of Xe of 21.0 and 32.1 eV, compared to 31.6 and 45.7 eV for K. Further reasons for the low ignition capability of K$_{8}$ are its long incubation time and its surface location. Delaying the initial ionization of K artificially until $t = -100$ fs (the instant when TI sets in for Xe), the detrimental effect of outer ionization of the seed electrons is reduced and the ignition probability increases to 0.6. Furthermore, the interior doping site of Xe brings the laser-driven cloud of quivering seed electrons in closer contact with the host cluster. The aspect of dopant location and dopant-He interatomic distance will be addressed later.

The average dopant charge state is considerably enhanced in case of ignition~\cite{Peltz:2011}, since EII as the almost exclusive ionization channel critically depends on the laser-driven nanoplasma electron cloud. This is demonstrated in Fig.~\ref{fig:Charging} b) for K$_{16}$ and Ca$_{8}$. In these examples the final dopant charges averaged only over trajectories with ignition (dashed lines) are by about 2 elementary charges higher than the corresponding values averaged over the entire trajectory set (solid lines).

Fig.~\ref{fig:Charging} e) exhibits the time-dependent average electron kinetic energies. Their time-dependent shape is determined by the instants of ignition. Trajectories without ignition almost do not contribute because of their small number of electrons.
The average electron kinetic energies obtained along single trajectories are in the range of $60$-$150$, $40$-$150$ and $140$-$150$ eV for K$_{16}$, Ca$_8$ and Xe$_8$ doped He$_{2171}$ droplets, respectively. The smaller trajectory set-averaged values for K$_{16}$ and Ca$_8$ as compared to Xe$_8$ doping are also due to the larger spread of ignition times. However, the reason for the higher average Xe charge [Fig.~\ref{fig:Charging} b)] is not the somewhat higher electron kinetic energies but the larger EII cross sections of Xe.

The outer ionization level in case of ignition can be derived from Fig.~\ref{fig:Charging} c) as the difference of the average He charge $q_{av}$ (solid line for Xe$_8$ and dashed lines for the K$_{16}$ and Ca$_8$ doped droplets) and the nanoplasma electron population $n_p$ per atom (dotted lines, obtained from the number of nanoplasma electrons within six droplet radii from the droplet center of mass). In all the three cases, the outer ionization amounts to about 0.5 elementary charges per atom.

A preliminary analysis shows that the main energy absorption takes place by the nanoplasma electrons during the avalanche EII. However, damping of the laser-driven nanoplasma electron oscillation is very strong as a large part of the absorbed energy is consumed by EII, such that a nanoplasma resonance is unlikely to be present during this phase. This changes only near the termination of avalanche ionization when damping becomes small and energy absorption is still considerable, apparently driving outer ionization during this time period.

A resonance also seems to be present during the incubation time, involving the seed electrons and a limited number of electrons set free from He ionization, when energy absorption is positive and damping is small. Although this small nanoplasma resonance does not carry weight in the total energy balance of the droplet, it might be important for the ignition process. Pristine He droplets for $I_M \geq 5\times 10^{14}$ Wcm$^{-2}$ (i.\,e., for intensities when TI rates at He become notable) show the same behavior: (i) resonance as long as only a limited number of He atoms is ionized, (ii) strong damping during avalanche EII and (iii) a second resonance with strong energy absorption towards the completion of the nanoplasma formation. Thus, the initial phase before the onset of avalanche EII resembles the experiment of Sch{\"u}tte et al.~\cite{Schuette:2016} where the seed electrons are generated by an XUV laser pulse. At lower pulse peak intensities (e.\,g., $I_M = 10^{13}$ Wcm$^{-2}$), avalanche ionization is not complete during the laser pulse, so that the only nanoplasma resonance occurs during the incubation time. These energy absorption and nanoplasma resonance phenomena as well as their implications for the ignition mechanism require further investigations. In this context it will also be of interest to characterize the damping of the nanoplasma in terms of the quality factor of damped oscillators.

\begin{figure}[hbt]
\center
\includegraphics[width=0.65\columnwidth]{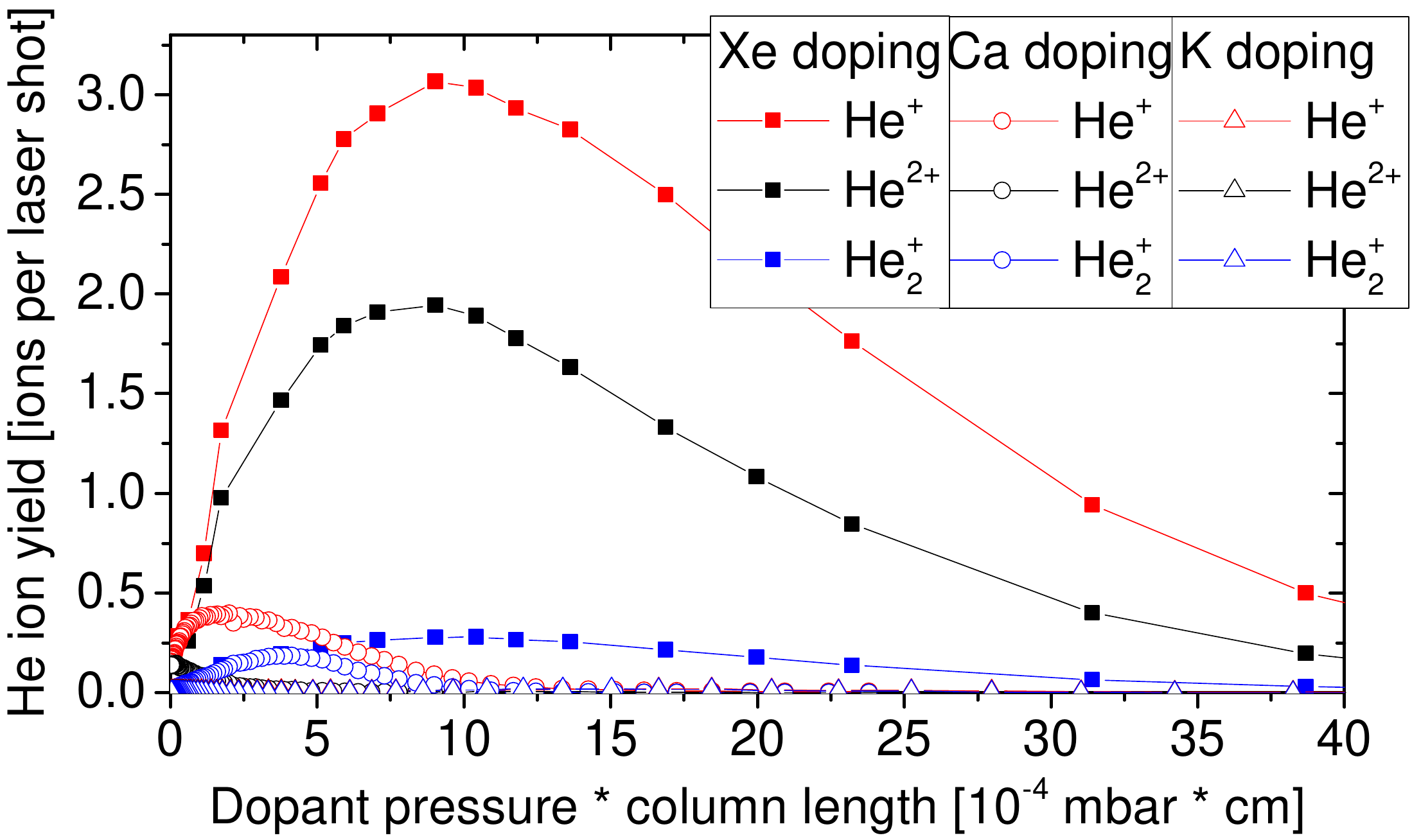} \caption{\label{fig:DopingCurve} Experimental He ion yields as a function of the vapor pressure of K, Ca, and Xe dopants multiplied by the length of the doping region (1~cm vapor cell for K, Ca, 35~cm vacuum chamber for Xe).}
\end{figure}

The efficiency of igniting a He nanoplasma is manifested by the appearance of He ion signals. Fig.~\ref{fig:DopingCurve} shows the experimental yield of He$^+$, He$^{2+}$, and He$_2^{+}$ ions recorded as a function of the vapor pressure of dopants. The latter is adjusted by controlling the temperature of the heated crucible in the case of K and Ca and by leaking Xe into the doping chamber using a dosing valve. The conspicuous result is that by far the highest He ion yields are obtained when doping with Xe, whereas doping with Ca and K provides lower He ion yields by about one and two orders of magnitude, respectively. When increasing the doping pressure starting from zero, the He ion yields first rise due to enhanced efficiency of the dopant-induced ignition process. The diminishing of ion yields for high doping pressures is a consequence of massive droplet beam depletion due to the release of binding energy when dopant atoms aggregate into clusters inside the droplets as well as scattering of the droplets away from the beam axis, as mentioned in Sec.~\ref{sec:experiment}. Effectively both density and size of the He droplets in the laser interaction region are thus reduced.

\begin{figure}[hbt]
\center
\includegraphics[width=1.0\columnwidth]{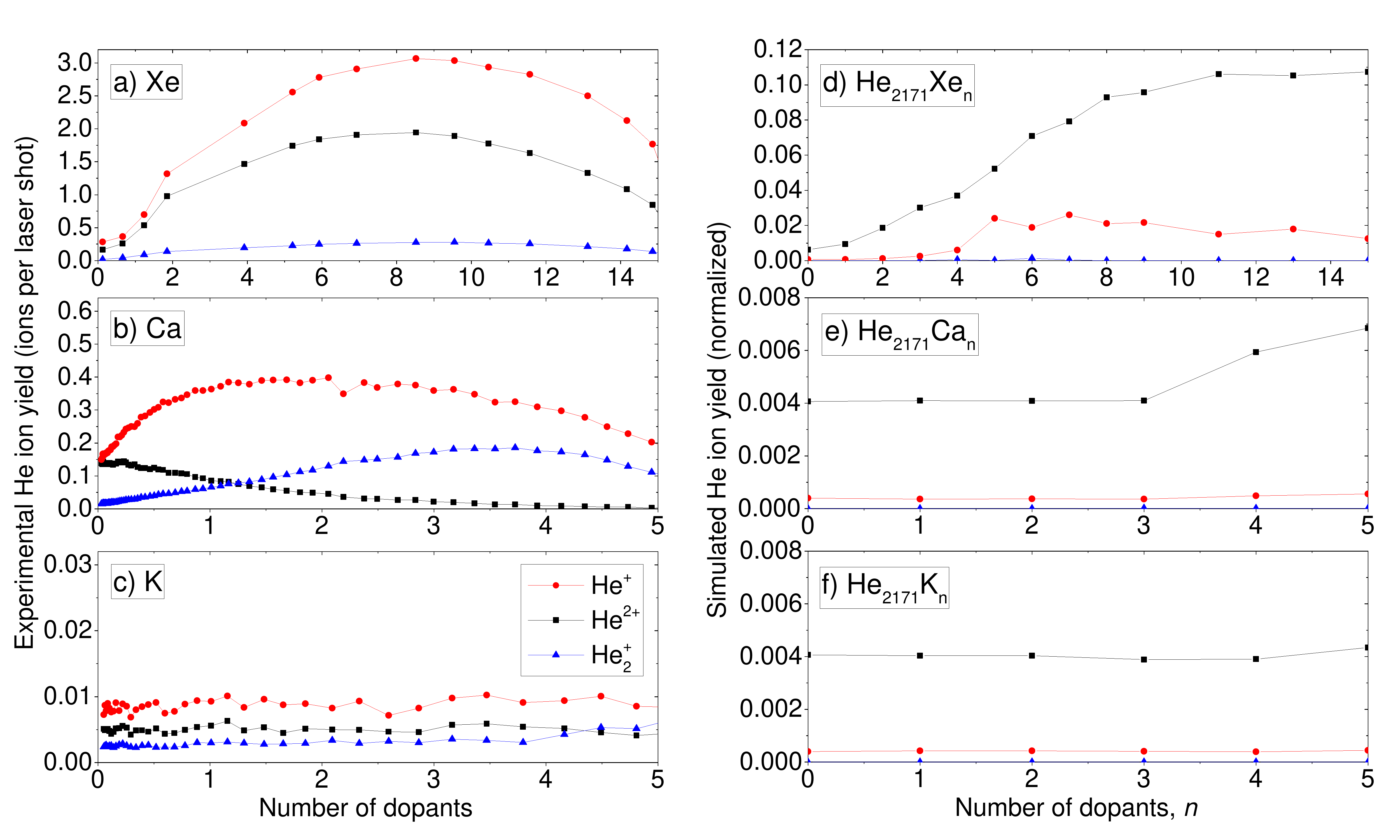} \caption{\label{fig:ExpSim} a)-c) Experimental and d)-f) simulated yields of He ions generated by He nanoplasma ignition induced by multiple dopants of the species K, Ca, and Xe. The simulated He ion counts in proportion to the total number of He atoms are averaged over the focus volume of the laser beam. In case of surface doping with K and Ca, the signals are also averaged over the parallel and perpendicular orientations of the dopant-droplet axis relative to the laser polarization.}
\end{figure}

In an attempt to directly compare the experimental results with the simulation, the experimental data of Fig.~\ref{fig:DopingCurve} are represented on different $x$ and $y$ scales in Fig.~\ref{fig:ExpSim} a)-c). The rescaling of dopant pressure to the number of dopant atoms relies on the detailed simulation of the doping process~\cite{Buenermann:2011}. The measured yields of He$^{+}$ and He$^{2+}$ ions as a function of the number of dopant atoms picked up on average by one droplet attains the highest values for Xe at about 13 dopant atoms. In contrast, the highest He$^{+}$ signal observed for K-doping stays below that for Xe-doping by factor $6\times 10^{-3}$. When doping with Ca atoms, the He$^{+}$ ion yield comes close to the one obtained for Xe-doping at low doping numbers ($n_{\mathrm{Xe}}\le $6), but continuously falls off as the doping level is increased. This diminishing of ion yields is primarily a consequence of massive droplet beam depletion by the large binding energy released when Ca clusters aggregate inside the droplets. Accordingly, the yield of He$^{2+}$ ions sharply drops upon doping only a few ($n_{\mathrm{Ca}}=2$-$3$) Ca atoms. The significant fraction of He$_2^+$ observed for $n_{\mathrm{Ca}}=3$-$6$ points at incomplete cluster ionization which is followed by dimer formation out of He$^+$ surrounded by neutral He atoms.

A complete simulation of the ion signals requires (i) the averaging over all intensities in the focal
volume which contribute to the He ion signal, (ii) the averaging over the size distribution of
doped droplets, (iii) the dopant size distribution, and, for surface dopant states, (iv) the averaging
over the orientation of the dopant-droplet axis relative to the laser polarization. For the simulated
He ion signals, Figs.~\ref{fig:ExpSim} d)-f), we have carried out only points (i) and (iv). Therefore, we cannot expect quantitative agreement between experiment and theory. As a consequence of the missing dopant size averaging, the abscissa of Fig.~\ref{fig:ExpSim} d)-f) represents a fixed number $n$ of dopant atoms, whereas each value of the abscissa of the experimental signals, Fig.~\ref{fig:ExpSim} a)-c), is the average number of dopant atoms in a distribution. 

\begin{figure}[hbt]
\center
\includegraphics[width=1.0\columnwidth]{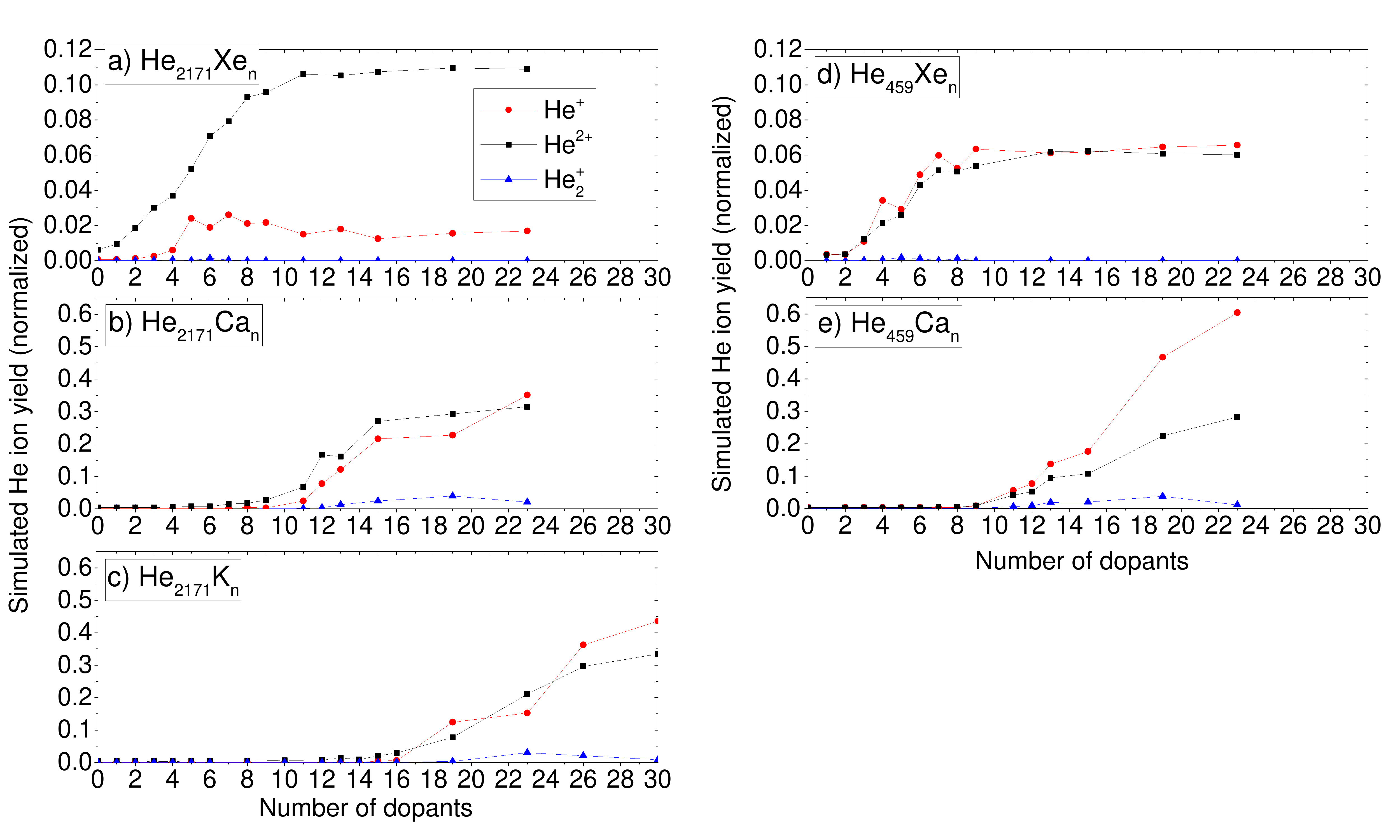} \caption{Simulated yields of He ions as a function of the dopant species for two different sizes of the He droplets, He$_{2171}$ (left column) and He$_{459}$ (right column). The data are averaged over the intensity distribution of the laser focus volume and over orientations of the K and Ca-doped droplets with respect to the laser polarization. \label{fig:TheoSignalsSizes}}
\end{figure}

The simulated results qualitatively reproduce the experimental He ion yields for the dopant sequence Xe $>$ Ca $>$ K in the regime of weak doping where the detrimental effects of droplet evaporation are nearly negligible. By far the largest He ion yields are obtained for Xe doping. For Ca and K doping, the small signal intensities stem from intensities $I\geq 5\times 10^{14}$ Wcm$^{-2}$ for which the droplet ignites by itself because of TI of He. Only for $n\geq 4$ for Ca and $n\geq 5$ for K, a slight signal increase occurs, when the dopants are able to induce ignition at the next lower intensity, $I = 2\times 10^{14}$ Wcm$^{-2}$, at which the focal volume is sampled. The experimental signal increase already for a small number of dopant atoms is caused by the admixture of signal intensities from larger dopants in the dopant size distribution and therefore cannot be reproduced by the simulations. Likewise, the decrease of the experimental signal for larger dopants because of droplet evaporation cannot be accounted for. The order of
the He$^+$ and He$^{2+}$ signal intensities compared to the experimental signal is reversed which we
partly attribute to the averaging over the broad He droplet size distribution. Simulations for the
smaller He$_{459}$ doped droplets as a sample of the droplet size distribution indeed show that the He
ion abundance considerably shifts towards singly charged He$^+$, as shown in Fig.~\ref{fig:TheoSignalsSizes} d) and e).

Since simulations are not limited by droplet beam depletion, we can study the hypothetical situation of attaching larger ($>10$) K and Ca dopant clusters to the He droplets. For these larger dopant clusters we find strongly enhanced He ion yields even for doping with Ca and K, see Fig.~\ref{fig:TheoSignalsSizes}. The larger number of dopant atoms supply enough seed ionizations for He ignition at low laser intensities which make up the largest part of the laser focus volume. 
Note that, as a consequence of focal averaging, for doping with $\gtrsim 11$ Ca or $\gtrsim 19$ K atoms which make the low intensities in the periphery of the focus volume available for ignition, the yield of the He ions even exceeds the maximum yield reached for Xe doping. The largest considered dopants K$_{30}$ and Ca$_{23}$ induce partial ignition already below $8\times 10^{12}$ Wcm$^{-2}$. The contribution of these very low intensities is on the order of 10\% and is neglected here. Since TI of Xe requires $I\gtrsim 5\times 10^{13}$ Wcm$^{-2}$, lower intensities remain unaccessible even for larger Xe dopants.

The systems considered in this work are dissimilar in various respects, given by the experimental boundary conditions: ionization energies and locations inside or at the droplet surface are different. In the experiment, droplet beam depletion upon cluster aggregation is a further dopant-specific limitation. What are the crucial factors for the observed conspicuous species-dependence of doped He nanodroplet ignition? In the following we systematically study the ignition efficiency of dopants in terms of their specific properties.

\begin{figure}[hbt]
\center
\includegraphics[width=0.55\columnwidth]{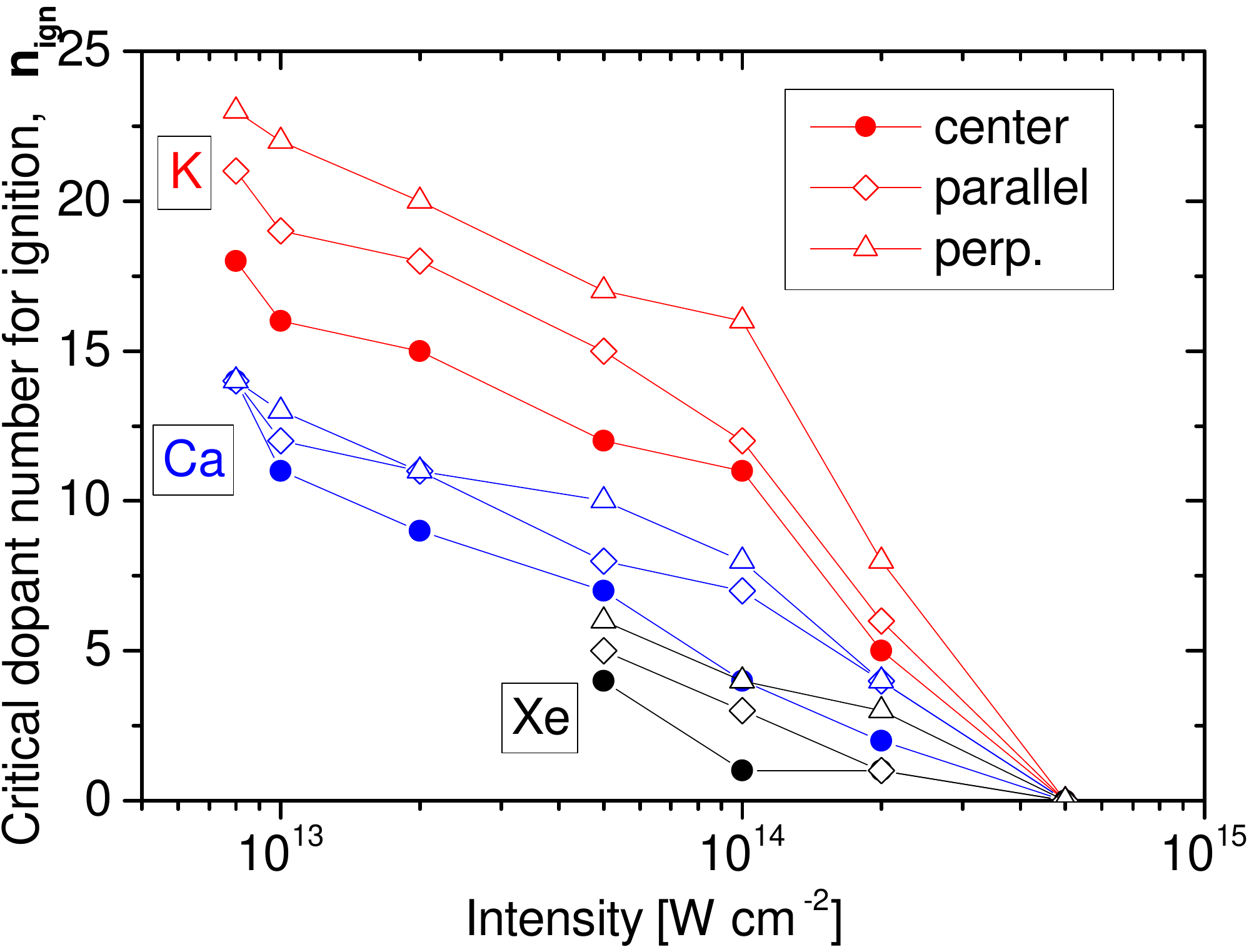} \caption{Simulated minimum numbers of K, Ca or Xe dopant atoms needed for ignition of a He$_{2171}$ droplet as a function of the (non-focally averaged) pulse peak intensity. Results are given for interior and surface doping, the latter for the parallel and perpendicular orientation of the cluster with respect to the laser polarization.}
\label{fig:Intens} 
\end{figure}

Fig.~\ref{fig:Intens} depicts the simulated minimum number $n_{\mathrm{ign}}$
of dopant atoms at which the ignition probability exceeds 10\%. The
variation of $n{}_{\mathrm{ign}}$ is shown as a function of the intensity $I$,
for interior and surface dopant states. The latter ones are distinguished
by parallel and perpendicular orientation of the dopant-droplet complex with
respect to the linear laser polarization. The data clearly show three
main trends in dopant-induced ignition: 

(i) A lower intensity can be compensated
to a large extent by larger dopant clusters. This is due to the larger
number of seed electrons available for EII and by the higher sum of
ion charges which assist EII by reducing the Coulomb barrier at He.
In addition, for a given species the number of seed ionizations per
dopant atom increases with the number of dopant atoms, as shown for
the K$_{8}$ and K$_{16}$ dopants (cf. Fig.~\ref{fig:Charging}).

(ii) Dopants which are easily multiply ionized (Xe, Ca) are considerably
favored, for the same reasons as in (i). 

(iii) Dopants residing in the droplet interior ignite the neighboring He atoms more efficiently, as the cloud of seed electrons quivering in the driving laser field has better contact with the He host droplet.
For all three dopants, a significantly larger number of dopants is needed for ignition of surface-bound dopant clusters at any laser intensity, where the parallel orientation is more favorable than the perpendicular one. 
Surface doping in parallel orientation typically requires 1-3 dopant atoms more to reach the same ignition efficiency as interior doping; the same gradation is found for surface doping between parallel and perpendicular orientation.

\begin{figure}[hbt]
\center
\includegraphics[width=0.6\columnwidth]{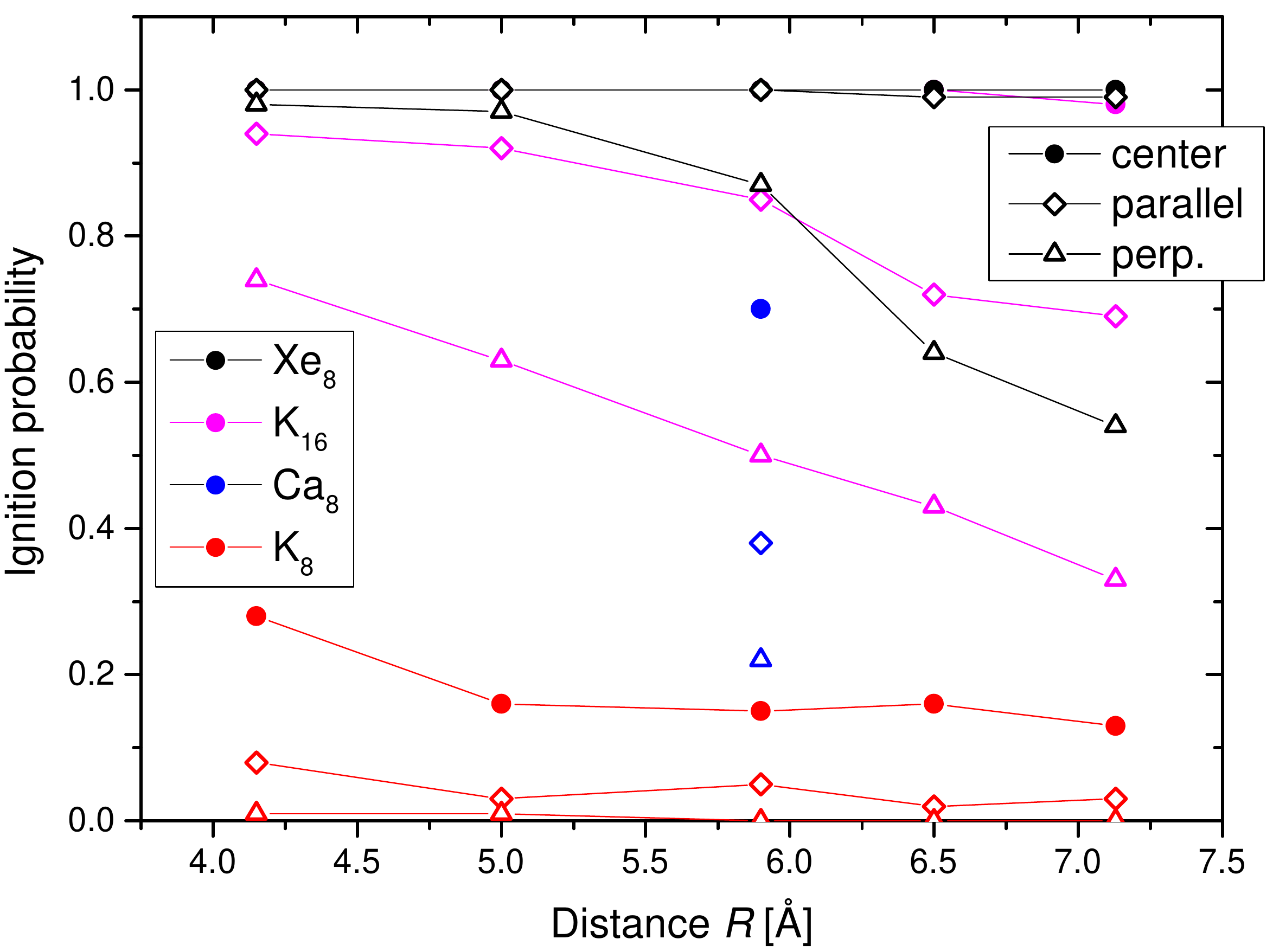} \caption{Dependence of the probability of igniting a He nanoplasma in a He droplet doped with various dopants as a function of the interatomic distance between dopant and He atoms. The peak laser intensity is $I=10^{14}$~W/cm$^2$ and the He droplet size is 2171 He atoms.} 
\label{fig:DistDep} 
\end{figure}

Another parameter which may severely impact the ignition efficiency of dopants attached to He droplets is the dopant-He interatomic distance as it affects the dopant-droplet contact strength as well. It is quite distinct for the three species under study and carries some uncertainty mainly due to unknown cluster size effects. Fig.~\ref{fig:DistDep} displays the dependence of the He ignition probability on the dopant-He distance at the interface between the dopant cluster and the He host matrix for the dopant samples of Fig.~\ref{fig:Charging}. 

Shown are the ignition probabilities in the interatomic distance range between the He-Xe ($4.15$ \AA) and the He-K distance ($7.13$ \AA) for interior as well as for surface doping in parallel and perpendicular orientation of the dopant-droplet axis relative to the laser polarization. While the general trend is, as expected, the decrease of the ignition probability with increasing He-dopant separation, Xe$_8$ in its interior doping state is so efficient that its ignition probability remains 1 in the entire considered distance range. In contrast, K$_8$ does not reach the ignition probability of Xe$_8$, even when it is brought to the droplet interior at the shorter He-Xe distance, confirming that geometrical effects alone cannot account for the larger ignition capability of Xe.

\section{Conclusions}

In conclusion, our investigations, theoretical and experimental, benchmark generic principles for the design of dopant clusters for efficient nanoplasma generation by examining the ability of K, Ca, and Xe dopants to ignite He droplets. The dopants' ability to trigger ignition of He droplets induced by EII at moderate laser intensities is determined (i) by the ability to provide initial seed electrons to drive electron-impact ionization of the He droplet, (ii) by the doping site -- interior or surface state -- and He-dopant distance which both determine the contact strength of the laser-driven quivering electron cloud with the He droplet, and (iii), as an experimental constraint, by the dopant's heat of cluster formation which induces partial droplet evaporation. While providing seed electrons remains a necessary condition for efficient dopant induced ignition, the ease of multiple, not just single, ionization of dopant atoms provides the key to droplet ignition. Surprisingly, low-lying first ionization energies of dopants which maximize the useful focal volume in an experiment are also detrimental to ignition since long incubation times favor outer ionization of seed electrons, as revealed in the case of K doping. 

Among the considered mono-elemental dopant clusters, Xe has the highest ignition efficiency, unifying the advantage of up to three seed ionizations per atom, occupying interior doping sites in the He droplet, and the lowest heat of dopant cluster formation. However, the first ionization energy imposes the limitation that seed ionization sets in only at $I\gtrsim 5\times 10^{13}$ W\,cm$^{-2}$. The possibility of optimizing the physico-chemical properties of the dopant clusters by mixing various species inside the same He droplet will be studied in a forthcoming work.

\subsection{Acknowledgments}
Stimulating discussions with Th. Fennel, M. Krishnamurthy and R. Gopal are gratefully acknowledged. This work is supported by the Deutsche Forschungsgemeinschaft in the frame of the Priority Programme "Quantum Dynamics in Tailored Intense Fields". The dissertation of B. Gr{\"u}ner is supported by scholarship funds from the State Graduate Funding Program of Baden-W{\"u}rttemberg. We are grateful to Ivan Infante for prepublication information on inner-shell ionization energies of K and Ca. The authors thank for computational and manpower support provided by IZO-SGI SGIker of UPV/EHU and European funding (EDRF and ESF).

\section*{References}

\providecommand{\newblock}{}

\end{document}